\pdfoutput=1

\documentclass{iucr}             

\usepackage{graphicx}
\usepackage{pythonhighlight}

\begin{document}                  



\title{Essential Scattering Applications for Everyone. Overview.}
\shorttitle{ESCAPE }


\author{Denis}{Korolkov}
\author{Stepan}{Rakhimov}

\aff{Email to: escape.app.net@gmail.com}









\maketitle                        


\begin{abstract}
\emph{ESCAPE} is a free python package and framework for creating applications for simulating and fitting of X-ray and neutron scattering data 
with current support for specular reflectivity,
polarized neutron reflectometry,
high resolution X-ray diffraction, small angle scattering with future support for off-specular scattering from structured samples with complicated morphology.
Utilizing current features of \emph{Jupyter} project (https://jupyter.org/),
it allows to create highly customized applications in the format of
\rm{notebooks}. These notebooks, being shared with other users, can be used directly or started as
web applications with graphical user interface.
This paper is a brief overview of the \emph{core} and \emph{scattering} packages providing description of the major features with code examples.
The following features make \emph{ESCAPE} different from other projects:
independent from scattering applications core, which provides access to models building blocks
like parameters, variables, functors, data objects, models and optimizers;
support of arithmetic operations and algebraic expressions
on parameters and functors, offering models with complex dependencies of parameters;
\emph{math} module with standard mathematical functors and special functors which perform numerical integration over variable or parameter,
supplying customization of intensity model;
simultaneous fit of several models, also for models with different dimensions.
Check our web  site  https://escape-app.net/ for further information.
\end{abstract}


\section{Introduction}

Due to continuously increasing number of industrial and scientific utilization of scattering methods
in condensed-matter (Orji et al., 2018) and soft-matter (M\"uller-Buschbaum, 2013) branches
it is important to provide for potential users
a flexible, highly customizable and relatively fast software for data analysis.

Scattering problems like small angle scattering, or grazing incidence small angle scattering,
specular reflectivity, diffraction, X-ray fluorescence etc., differ significantly not only in terms of theory and
experiment handling, but also in data preparation and visualization.

Many existing applications are dedicated for a particular sample type and particular
scattering method. As a result users have to deal with several data formats,
several optimization routines, several, sometimes not obvious, workflows.

Application of new data analytics such as big data handling, deep learning, artificial intelligence techniques
require easy integration of modelling and simulation routines to the existing frameworks and infrastructure (see report IEEE (2018)). 

Developing a software which successfully resolves all issues and satisfies advanced users and beginners
is a challenging task.

In this publication we give an overview for the Essential Scattering Applications for Everyone (\emph{ESCAPE}),
a private independent project.
\emph{ESCAPE} is compiled as a python package which, compared to other existing scattering applications 
(Lazzari (2002), Bressler et al. (2015), Pospelov et al. (2020)) provides a set of common
interfaces with the same look and feel for different scattering techniques fully integrated by default with the
Jupyter notebook environment.

The development of this software and its first early prototype dates back to 2012. The idea of the software concept came in author's mind
during High Data Rate Initiative sessions (https://www.pni-hdri.de).  
\emph{ESCAPE} is dedicated first of all to experienced physicists - scattering instrument or laboratory responsibles, people
who understand in details scattering techniques, data preparation and required scattering formalism for obtaining
successful results. With \emph{ESCAPE} they can prepare a set of customized applications (also with graphical user interface)
for different experimental setups and share these applications with less experienced users. After publication of results users
can share their experience with others simply publishing notebooks in our repository,
check our webpage https://escape-app.net for more details.
As a result everyone can use shared notebooks for educational purposes, training and planning of future experiments,
simulating experimental curves for new samples and playing with settings and parameters of models.
The present paper focuses on a general concept of \emph{ESCAPE} and core features.

\section{Overview}

\subsection{License and collaborative work}
\emph{ESCAPE} is released closed source and is free for non-commercial usage.
Free availability together with widely used Jupyter notebook (Kluyver et al, 2016) file format provides full verifiability of obtained results.
A clear, continuously developed python modules allows experienced users to adapt \emph{ESCAPE} to their own needs.

Examples of notebooks are published under GNU General Public License version 3. Users of \emph{ESCAPE} project are encouraged to
share their notebooks with published scientific results. The published notebooks give information about most often usage cases
of \emph{ESCAPE}, inducing authors for further improvement of available modules and resolving issues.

\subsection{Programming languages}

Different programming languages have been created for different purposes and in \emph{ESCAPE} we try to use
programming languages in the most effective way. The core of \emph{ESCAPE} is a template, header only library
written in C++, using standard C++17.
C++ language provides good performance and parallel computation. The core is fully independent of any other libraries
providing relatively small compiled size of the package.
Python bindings are developed with Cython. For most of the C++ classes in the core there is a wrapper Cython class.
This wrapper classes are visible for users and define the package functionality available for users.
Cython wrappers are split into several modules and compiled
for Linux, MacOS and Windows platforms as Python \emph{whl} installation archives. \emph{ESCAPE} has also \emph{escape.utils} package with useful specific modules and routines
to simplify some usage cases. This includes material database, based on \emph{periodictable}  package  and special commands for calling compact,
integrated with notebooks widgets to simplify certain routines.

As a result, for users who are familiar with Python itself and scientific Python packages like \emph{scipy} (Virtanen et al., 2020),
\emph{numpy} (van der Walt et al., 2011), \emph{matplotlib} (Hunter, 2007), etc., the learning curve will be shallow.
For others, it will be another good reason to learn one of the most popular platforms nowadays.

\subsection{Core package overview}

In order to describe a core functionality and the main philosophy of \emph{ESCAPE} package,
we provide in this section a simple example. Imagine that you have some experimental Small Angle Scattering Data
from silica particles of spherical shape. You would like to create a model and to fit these data.
Creation of a model in \emph{ESCAPE} starts with definition of parameters. A standard scattering form-factor of a spherical particle
is given by the following equation

\begin{equation}\label{eq:formula_sphere}
F(q)=3\frac{\sin(qR)-qR\cos(qR)}{(qR)^3}
\end{equation}

where $q$ is a length of scattering vector, $R$ is a particle radius.

The measured intensity is given as following

\begin{equation}\label{eq:formula_sphere_intensity}
I(q)= N\Delta\eta^2V^2|F|^2+B
\end{equation}

where $N$ is the number density of particles, $V=4/3\pi R^3$ is the volume of the particles,
$\Delta\eta$ - scattering length density contrast and $B$ - constant background.

We define parameters of the model, variable and intensity functor, i.e. function object, as following

\begin{python}
import escape as esc
import numpy as np
from escape.utils import show

R = esc.par("R", 7.5, units="nm")
C = esc.par("Contrast", 1.2, scale=1e-3,
             units="nm^-2")
B = esc.par("Background", 4, units="arbitr")
N = 1e5
q = esc.var("q")
V = 4/3*np.pi*esc.pow(R, 3)
F = 3*(esc.sin(q*R)-q*R*esc.cos(q*R))/
    (esc.pow(q*R, 3))
I = N*esc.pow(C*V*F, 2.0)+B
show(I, coordinates=np.arange(0, 4.0, 0.001),
     ylog=True,
     ylabel='I', xlabel='Q[1/nm]',
     title="Intensity: Spherical particle")
\end{python}

\begin{figure}
\includegraphics[width=\textwidth]{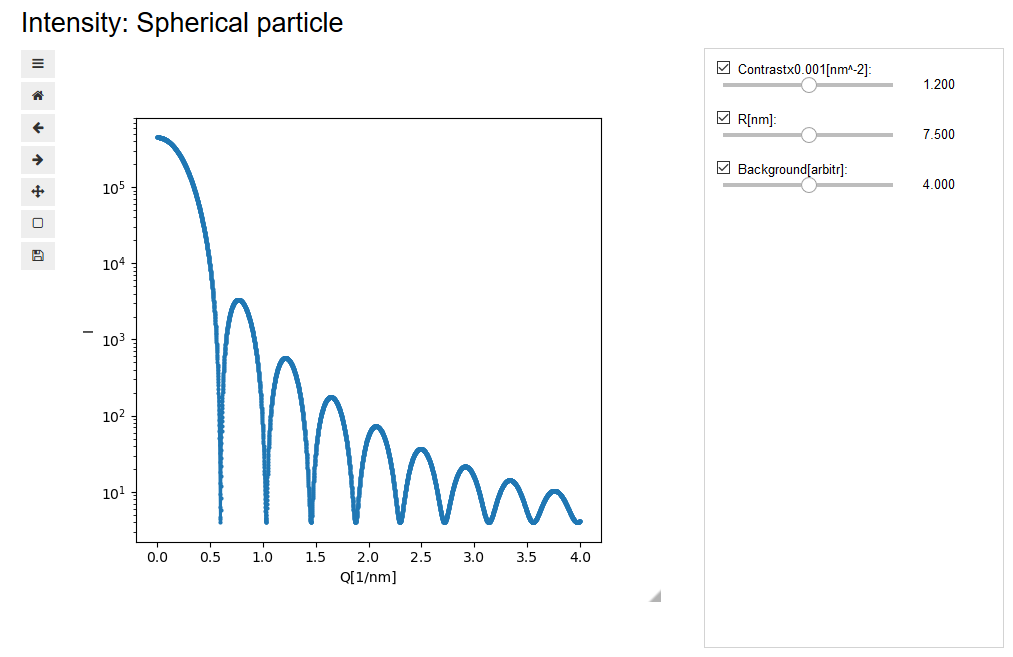}
\caption{Widget for the intensity functor.}
\label{fig:widget1}
\end{figure}

A result of this code is a widget (see Figure \ref{fig:widget1}), with
controls to change parameters values on the right side and resulting plot of intensity.

Next we open the experimental data using \emph{numpy.loadtxt} routine and create data object \emph{dobj}- container for
experimental data. Model object \emph{mdl} calculates simulated model curves and can compare
them with experimental data by means of the cost function. For the optimization we are going to use Levenberg-Marquardt optimizer (see section \ref{sec:optimizer}).

\begin{python}
x, y = np.loadtxt("example1.dat",
                  unpack=True)
dobj = esc.data("Data", x, y, copy=True)
mdl = esc.model("Model", I, dobj)
opt = esc.levmar("LM", mdl)
show(opt, ylog=True,
     ylabel='I', xlabel='Q[1/nm]',
     title="Intensity fitting")
\end{python}

\begin{figure}
\includegraphics[width=\textwidth]{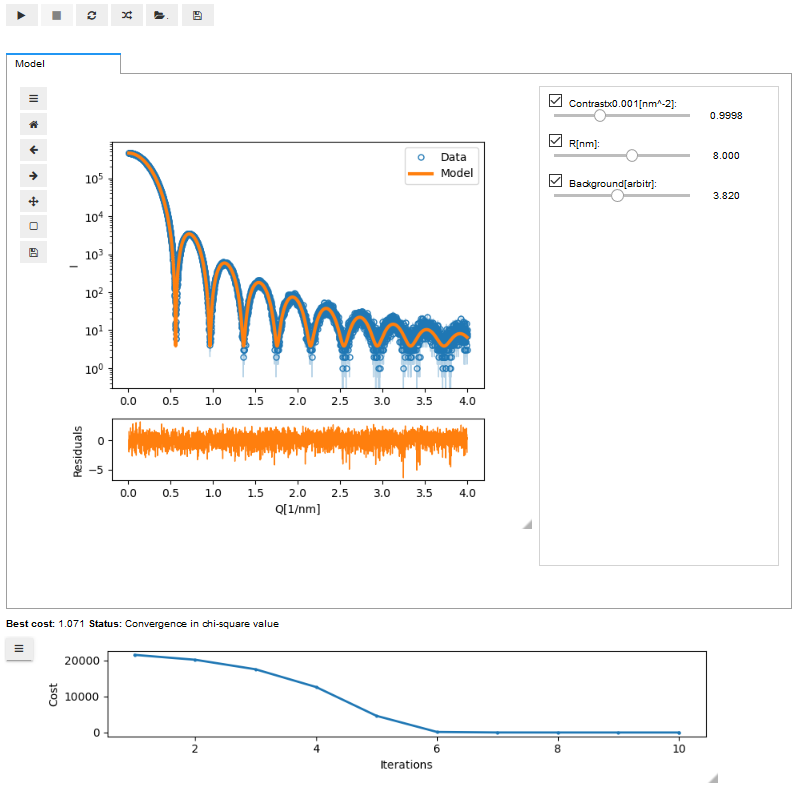}
\caption{Widget for the optimizer}
\label{fig:widget2}
\end{figure}

Resulted widget (see Figure \ref{fig:widget2}) provides the plot with a cost function history and can be used to start/interrupt optimization process
and save/load parameters values with estimated errors.

As it has been already mentioned, a user should not create such a notebook from scratch. For the beginning we have plenty of example notebooks for
all kind of available applications. 

\subsubsection{Parameters}

Internally \emph{ESCAPE} package operates with several parameter handlers of two sorts: independent parameters, which are recognized
for optimization and dependent parameters, which have, for example, a form of arithmetic expression and depend on parameters of the first type.

Practically both groups of parameters are hidden behind the same object type and user can use
a dependent parameter of any complexity where appropriate.
This gives a lot of freedom for definition of complex problems where independent parameters being a part
of complex expression are shared between different objects, like layers in multilayer samples, for example,
or even between different models, which can be fit simultaneously.

\subsubsection{Functors}

Functors are objects which behave like functions, i.e. they have a \emph{call} operator.
In \emph{ESCAPE} we tried to make functors definition as simple as writing a formula.
Functors support up to five variables, which is enough for scattering applications implemented in \emph{ESCAPE}. On request, the number of variables can be 
easily increased.

Functors in \emph{ESCAPE} can be defined as a mathematical expression, like spherical form-factor we presented above. 
More complicated functors, for calculation of intensities or averaging, etc. are implemented in C++ core
and created by user using special factory functions.
Functors support arithmetic operations which provides a flexible way of customizing
intensity calculation with specific diffuse scattering, background with scattering vector dependency,
illumination correction, resolution effects, etc.

Most of the functors in \emph{ESCAPE} return float values when being called, but additionally, there is a support for 
\emph{complex functors} which operates real type variables but return complex values.  

\subsubsection{Math}
Parameters and functors in \emph{ESCAPE} support arithmetic operations and mathematical functions. Additionally
to standard mathematical functions like $\exp$, $\log$, $\mathrm{pow}$, $\sin$, $\cos$, etc., there are also special functors
implemented which perform numerical integration over variable of an integrated functor or its parameter. This gives a possibility
for integrating of intensity over resolution function, or averaging integration over certain parameter and its distribution.
In the examples available online, we provide notebooks which demonstrate usage of such integral functors,
over parameter and variable, for example, for the intensity simulation from bimodal silica nanoparticles
with size distribution, like in the paper Bre{\ss}ler et al.(2015). Integral functors support two integration methods: with fixed number of quadrature points and adaptive.
The characteristic width of distribution function in the case of averaging integration can be defined as a functor of integration variable,
which allows to perform integration for the cases of instrumental resolution function with varying full width of a half maximum.

\subsubsection{Data}

Data object is a container which prepares experimental data for optimization.
Currently there is only generic implementation which can handle provided arrays of
coordinates, intensities and errors. Automatic reading of experimental data from
a file is specific for every instrumental setup and is not supported currently.
If experimental file has a format of a text tabulated data, users can try to open it using \emph{numpy}
python library.
Errors of intensities can be provided as an array or will be
calculated from intensity values according to Poisson statistics as $\sqrt{I}$ by default.
The correctness of provided data errors defines the correctness of estimated parameters errors, made after fitting. 
Data object also supports masking of experimental data, which allows to ignore
points with detector failures, cosmic rays, primary beam and other parasitic features.

\subsubsection{Model}

The main task of the model object is to make a quantitative comparison of results
returned by the functor, i.e. simulation curve, with experimental data returned by the data object.
This comparison is made by means of residuals and well-known  $\chi^2$ cost function which
is given by

\begin{equation}
\label{eq:cost}
\chi^2=\frac{1}{N-M}\sum_{i=1}^{N}\left[\frac{I_{\mathrm{exp}(q_i)}-
I_{\mathrm{mod}(q_i)}}{\sigma_{\mathrm{exp}(q_i)}}\right]^2
\end{equation}

The intensities in equation \ref{eq:cost} can be scaled, which is important if there is a power low of intensity values or
there is a strong peak in experimental data which is several magnitudes stronger than intensity values in its neighbourhood.   
Model object provides several scaling options for residuals:
 logarithmic scale of intensity, and scaling of intensity with the scattering vector of power
two or four.   
In future we plan to add a support for fully custom, user defined scaling functors.
The residuals scaling is used for the calculation of cost function during the fit, 
estimation of errors is done always with standard cost function without scaling. 

\subsubsection{Optimizer}
\label{sec:optimizer}
Among most of existing optimizers we have chosen Levenberg-Marquardt - well-known
damped least-squares optimization method suitable for most of the scattering problems and
Differential Evolution (DE) method, stochastic method, which is very simple to implement and to extend.

The Levenberg-Marquardt
optimization algorithm (Levenberg, 1944; Marquardt, 1963) is a least-squares method which interpolates between Gauss-Newton method and gradient-
descent and is known to converge fast to the local minimum of the cost function.

Scattering problems, especially for multilayer, could have a large number of parameters, and depending on quality of
experimental data it could lead to several cost function minima in the search domain. If constrains of parameters are too wide
and initial model curve is too far away from experimental curve, this can lead to instabilities in found solution.

To prevent possible instabilities it
is required to start iterative LMA optimization when the model intensity is close to
experimental values. This is quite often considered as the main disadvantage of LMA optimization method.

In contrast differential evolution method is a meta-heuristic method, it doesn't require
optimization problem being differentiable and depending on settings can search
larger spaces of candidate solutions comparing to LMA.

The method of differential evolution was developed by Price and Storn (see Storn et al., 1997).
To our knowledge, the first usage of this algorithm for scattering applications is described in Ref. 
Wormington et al. (1999).
The idea of this algorithm is the following:
initially a population of points $p$ in $m$-dimensional space is generated and
cost function is evaluated for each point. Then at each iteration for each point
$p_i$ three different points $p_a$, $p_b$ and $p_c$ are randomly chosen from
the population. The next step is a mutation - a new point $p_n$ is constructed
from these three points as following $p_n=F(p_b-p_c)+p_a$ with a probability
of crossover setting $Cr$. If a cost function for a new point $p_n$ has a better
value than the currently chosen $p_i$, than it replaces $p_i$.
If not than $p_i$ remains. Thus on each iteration we obtain a population with
points which are better or as good as previous. The process is repeated
until maximum number of iterations is reached.

We have slightly extended our implementation of DE optimizer.
There are two polishing scenarios for parameters: final polishing after DE
optimization and candidate polishing, when point candidate is
succesfully chosen to replace $p_i$ in population set. Both are done with LMA
optimizer and for both one can set maximum iteration number. In the latter case
polishing with only few iterations could improve convergence and reduce number
of function evaluations in our tests. This also resolves partly the main
disadvantage of DE compared to LMA - a slower convergence and quite
high number of function evaluations. DE in most cases is not sensitive
to initial solution and is not easily trapped by local minimum like LMA.
But still there is no guarantee that DE will find global minimum.

\subsubsection{Machine learning. Regressor}

A solution for typical inverse problems like scattering applications,
can be found not only by means of optimization, i.e. comparison between measured and simulated intensity.
Encouraged by work of Glorieux et al. (2001), we present here a \emph{regressor}
to solve inverse problems by means of neural network regression algorithms. The corresponding class plays a role of an interface for
any regressor written in python, which satisfies certain functionality. As a basis for this interface we took 
\emph{MLPRegressor} class from \emph{sklearn.neural\_network} package. Below is an example of 
how to use regressor with an existing model.

\begin{python}
from sklearn.neural_network import MLPRegressor 
rgi = MLPRegressor(max_iter=1000,
                   solver="adam",
				   activation="relu")

def noisify(X):
    X = np.asarray(X)
    X = np.random.poisson(X)
    return X

def normalize(X):
    X = np.asarray(X)
    X = np.log10(X*N+1)
    X = X/np.max(X)
    return X

rg = esc.regressor("Regressor",
                   mobj, 1500, 150,
				   impl=rgi,
				   norm_method=normalize,
                   noisify_method=noisify)
show(rg, ylog=True)
\end{python}

\begin{figure}
\includegraphics[width=\textwidth]{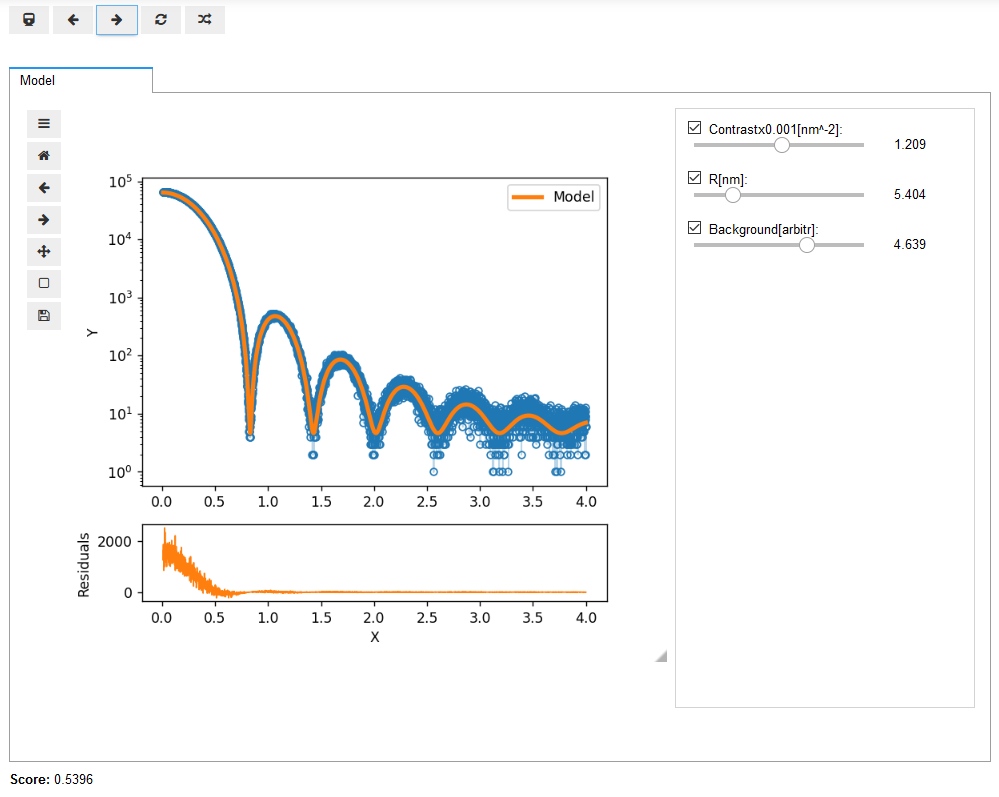}
\caption{Widget for the neural networks regressor object.}
\label{fig:regressor}
\end{figure}

We create first the \emph{MLPRegressor} object from \emph{sklearn.neural\_network} with \emph{adam} solver, which refers to stochastic gradient-based optimizer,
and \emph{relu} activation function for the hidden layer, which returns $f(x)=max(0,\;x)$. 
A complete list of all arguments and their possible values can be found in the official documentation for the \emph{sklearn.neural\_network} package. 
Next we define two methods, \emph{noisify} and \emph{normalize} which modify the generated data by adding Poisson noise, scale and normalize the intensity values.
Particularly, the \emph{noisify} method adds the Poisson noise to the data and  \emph{normalize} method performs logarithmic scaling and normalization to the maximum value.
The created \emph{MLPRegressor} instance together with the data modification methods are the input parameters of \emph{escape.regressor} factory function.
Before regressor will be able to make any prediction it must be trained first. The request for the training can be done in the widget
(see Figure \ref{fig:regressor}) by pressing corresponding button or by calling a corresponding \emph{train} function of the created instance. 
Before training regressor wrapper generates a certain number of experimental data sets, $1500$ in our case, by changing model parameters randomly 
and applies \emph{noisify} and \emph{normalize} methods to the data arrays,  
then it starts the training procedure of the \emph{MLPRegressor} instance on the generated and modified dataset.
Together with the training data set regressor
generates the test data set to calculate the training error. The regressor widget allows user to check prediction results for the test data set. 
After training user can perform prediction of parameters for real experimental data by calling regressor object as \pyth{rg(dobj)}, where \emph{dobj} is a data object, containing
experimental data arrays.
The data normalization method provided during creation of regressor instance will also be applied to the experimental data.

Prediction quality is not perfect and depends on the number of training datasets and number of free parameters in the model. 
Overall, our conclusions are not far away from the conclusions
presented in Glorieux et al.(2001), where neural networks regressor was applied to HRXRD model with gradient layer.
The number of sublayers used to describe a gradient of SLD was also one of the justified parameters during regression.
\emph{ESCAPE} currently doesn't allow that and predicts only parameters of the model, i.e. quantitative model characteristic.
But, certainly, the implementation of prediction of qualitative model settings,
such as number of layers, gradient function, roughness model, etc., is of our great interest.

The regressor doesn't guarantee perfect result, but can be used for preliminary estimation of model parameters before fitting.
It can reduce number of optimizer iterations needed for convergence of optimizer to the minimum of the cost function and can be effectively used
for the batch processing, i.e. for the large number of experimental data obtained from the same type of sample, which is a typical industrial application.

\subsection{Scattering package overview}
The \emph{scattering} package contains components for material and sample description as well as functors for
calculation of intensity for different scattering applications. Currently sample description objects are dedicated for layered
samples with homogeneous layers and layers with gradients of
material parameters, like scattering length density, strain, etc.
The support for structured samples comes together with \emph{potential} module, which is experimental
currently.

As we have already mentioned, all scattering problems in \emph{ESCAPE} are implemented in terms of functors.
The variables for this functors are provided by user and for scattering applications these variables
are always reciprocal space coordinates, like components of scattering vector or scattering vector length.

\emph{ESCAPE} doesn't operate entities like scans types or description of experimental setup, like detectors or slits.

These entities are very specific to experimental setups and in most cases users can easily convert
real-space parameters of performed experiment,
like forthcoming and outgoing angles of the primary and scattered beam
to corresponding reciprocal space coordinates of scattering events on the detector. 
Below we give a brief overview of \emph{scattering} package modules.

\subsubsection{Materials}

The speed of propagation of plane waves in the medium is slower than in vacuum. The index of refraction is a commonly used
characteristic of material, which is equal to ratio between the speed of wave in vacuum and inside a medium.
For a homogeneous material it is defined as following

\begin{equation}\label{eq:refraction}
n=1-\delta+i\beta
\end{equation}

Parameters $\delta$ and $\beta$ are very well-known in optics and usually are used to characterize a material
in scientific software packages for X-ray scattering. Both parameters depend on composition of medium and wavelength of incident wave.

In \emph{ESCAPE} we use  Scattering Length Density (SLD), both real and imaginary (absorption) parts, as default fit parameters.
SLD is more popular in neutron scattering community and conversion between terms of refractive index and SLD is the following:

\begin{equation}\label{eq:refraction}
\delta=\frac{\lambda^2}{2\pi}\mathrm{Re}(SLD);\quad\beta=\frac{\lambda^2}{2\pi}\mathrm{Im}(SLD)
\end{equation}

Another approach, very popular in commercial applications is to fit a mass density instead of refractive index terms.
This approach is also available in \emph{ESCAPE} using the material database package (see \emph{escape.utils.mdb}).
A composition of material can be provided as a chemical formula, in the same way as in \emph{periodictable} package, which is used by the module.
The resulted \emph{material} instance depends on mass density parameter provided by user.
Both SLDs, real and imaginary part are calculated from the mass density. It reduces the number of fit parameters, but from our
experience work good only for the cases of perfect quality samples, when composition of an investigated sample is known.

\emph{ESCAPE} allows to work with both approaches and to apply them simultaneously for creation of material objects.

Below we define two \emph{material} objects for Fe and Si ignoring for simplicity their absorption scattering length density parameters

\begin{python}
Fe_sld=esc.par("Fe SLD", 8.024,
               scale=1e-4, units="1/nm^2")
Si_sld=esc.par("Si SLD", 2.074,
               scale=1e-4, units="1/nm^2")
Fe = esc.amorphous("Fe", Fe_sld, 0.0)
Si = esc.amorphous("Si", Si_sld, 0.0)
\end{python}

\subsubsection{Layered sample}

The modelling of intensity cannot be done without sample description. For this purposes \emph{ESCAPE} operates several objects of the
following types:  \emph{layer}, \emph{layerstack} and \emph{multilayer}. The \emph{layer} instance is used for description of a
single layer and requires \emph{material}, thickness parameter and root mean square roughness parameter. In the code listing below we create
three layer objects forming trilayer Fe/Si/Fe with zero roughness. The Fe layers share the same material object and, thus, there SLDs parameters are also shared.

\begin{python}
thknFe1 = esc.par("Fe Thkn 1", 10, units="nm")
thknFe2 = esc.par("Fe Thkn 2", 10, units="nm")
thknSi = esc.par("Si Thkn", 1, units="nm")
#zero roughness
Fe_layer1 = esc.layer("Layer: Fe 1", Fe,
                      thknFe1, 0.0)
Fe_layer2 = esc.layer("Layer: Fe 2", Fe,
                      thknFe2, 0.0)
Si_layer = esc.substrate("Substrate: Si", Si, 0.0)
\end{python}

If multilayered sample consists of stack of several layers repeated many times, \emph{layerstack} object should be used which
takes number of repeats of the stack in the multilayer sample.

Stacks and layers should be added to \emph{multilayer} object as following
\begin{python}
#substrate material, with constant SLD
Si_sub = esc.amorphous("Si", 2.074e-4, 0.0)
air = esc.air("Air")
sub = esc.substrate("Substrate: Si", Si_sub, 0.0)

sample = esc.multilayer("Fe/Si/Fe", air, sub)
sample.add(Fe_layer1)
sample.add(Si_layer)
sample.add(Fe_layer2)

show(sample, xlabel="Z[nm]",
     ylabel="SLDRe[1/nm^2]", yaxis="sld0re")
\end{python}

The substrate material we left with constant standard value of SLD. Background (substrate) and foreground (air) layers both have zero roughness and
no fit parameters. When being added new layers are always inserted between the substrate and the last layer. Indexing
of layers starts from the first layer after foreground down to the substrate.
 The last command in the code listing will show the widget with a plot for real part of scattering length density profile.

\subsubsection{Specular reflectivity}

Specularly reflected intensity from a multilayer with rough interfaces can be calculated in terms of dynamical, semi-kinematic
and kinematic approximations of scattering theory. A good review of these approximations with a full list of references and their application to 
specular and off-specular reflectivity can be found in the thesis of Petr Mikulik, 1997.
For the dynamical theory there are two formalisms exist: transfer matrix formalism and Parratt formalism.
The latter is the default formalism for the modelling of specular reflectivity in \emph{ESCAPE}. 

Semi-kinematic approximation takes into account only single reflection in every layer and, thus, gives wrong result
near the angle of total reflection.
The kinematic approximation doesn't take into account absorption of incident and reflected wave inside the layers and fail for thick samples also at higher angles.

Functor object which calculates specular reflectivity using transfer matrix formalism is created as following
\begin{python}
Qz=esc.var("qz")
R = esc.specrefl("Specrefl", Qz,
                 sample,
				 formalism="matrix")
\end{python}


\subsubsection{Polarized neutron reflectivity}

Polarized-neutron specular reflectometry (PNR) allows to measure depth-resolved magnetization in flat films
with characteristic thicknesses from 2 to 5000 \AA. This method has been widely used to study homogeneous
and heterogeneous magnetic films, as well as superconductors.
The off-specular reflectivity scans allow to characterize lateral magnetic structures.

Implementation of Polarized Neutron Reflectivity functor is based on the following publications:
R\"{u}hm et al. (1999), Kentzinger et al.(2008).

In \emph{ESCAPE}, we provide a functor which accepts as input parameters efficiencies for polarization and analysis.
The sign of this parameter defines a direction of polarization.
Reflectivity functors which includes all four channels $++$, $--$, $+-$, $-+$ for simultaneous fit can be defined as following

\begin{python}
Qz = esc.var("Qz")

#polarization efficiency
poleffi = esc.par("Poleff_i", 1)
polefff = esc.par("Poleff_f", 1)

#flipper efficiency
flipeffi = esc.par("Flipeff_i", 1)
flipefff = esc.par("Flipeff_f", 1)

Rpp = esc.pnrspec("R++", Qz, sample,
                  poleffi, polefff)
Rpm = esc.pnrspec("R+-", Qz, sample,
                  poleffi, -polefff*flipefff)
Rmp = esc.pnrspec("R-+", Qz, sample,
                  -poleffi*flipeffi, polefff)
Rmm = esc.pnrspec("R--", Qz, sample,
                  -poleffi*flipeffi, 
				  -polefff*flipefff)
\end{python}

where \emph{sample} is a layered sample object. This functors can be added to model objects and fit later simultaneously.
This approach allows to avoid additional experimental data correction.
If efficiency is known, values of corresponding parameters might be corrected manually and fixed or set as constant values to disable their optimization.

\subsubsection{High resolution X-ray diffraction}

Next functor which we would like to present is for simulating dynamical x-ray diffraction from strained crystals, multilayers, and superlattices.
Its implementation is based on publication of Stepanov et al. (1998) and is based on two-beam approximation.
This functor provides solution only for scans near a single diffraction peak.
There are implementations for two approximations available: 4x4 matrix algorithm - valid for a full range of outgoing angles,
including low angles,
where reflection amplitudes are strong; 2x2 matrix algorithm - valid for high angles, where reflection amplitudes are negligible.
The first one is normally used for modelling of diffraction at grazing incidence and the second one for the conventional diffraction scans.

The functor can simulate the following structural properties in multilayer: characteristic of material, like SLDs, normal lattice strains,
interface roughness.

For the simulation of lateral strain effects additional scripting is necessary, an example notebook is in preparation.
The current implementation cannot be used for scans far away from the Bragg peaks and for the scans which include several diffraction peaks.
Currently there is no implementation of kinematic and semi-kinematic approximations.
We plan to add them in future versions together with support for structured samples.

\subsubsection{Structured samples. Potential}
\label{sec:potential}
The \emph{potential} module is a basis for description of structured samples.
It is currently experimental and has been probed only for Small Angle Scattering applications,
but certainly will be developed further and used for Diffraction, Grazing Incidence, 
and Off-specular Reflectivity from structured samples. 
The advantage of this module is a possibility to create complicated structures by means of arithmetic operations on potentials.
After releasing, users will be able to build lithographically structured samples with a complexity comparable with samples presented in 
Sunday et al.(2019).

Imagine that we would like to get the potential object of a structure unit consisting of two rectangular fins,
located at distance $D$ from each other.
A regular lattice of such fins can be a part of semiconductor device used in electronics.
So, it is not far away from real application. Each fin has a width of $w_x=25\, nm$, and consist of silicon core covered
with $2.5\, nm$ hafnium oxide. We create first two materials objects as following

\begin{python}
qx, qy, qz = esc.var(['qx', 'qy', 'qz'])
si = esc.amorphous("Si", 20.071e-6,
                   -0.458e-6)
hfo2 = esc.amorphous("HfO2", 63.976e-6,
                     -4.178e-6)
\end{python}

Next we create necessary parameters for silicon core and hafnium oxide shape.

\begin{python}
wx_f = esc.par("wx f", 25, units="nm")
wy_f = esc.par("wy f", 500, units="nm")
wz_f = esc.par("wz f", 100, units="nm")

wx_c = esc.par("wx c", 20, units="nm")
wy_c = esc.par("wy c", 500, units="nm")
wz_c = esc.par("wz c", 97.5, units="nm")
D= esc.par("D", 75, units="nm")
\end{python}

And now we are ready to create potential objects. Calculation of potential of structured samples is easy to
define in terms of the shape function $\Omega(\mathbf{r})$. For the fin structure we are trying to describe
the equation for the potential can be given as following

\begin{equation}\label{eq:fin_potential}
V(\mathbf{r})=V_{o}\Omega_{f}(\mathbf{r}) - V_{o}\Omega_{c}(\mathbf{r}) + V_{c}\Omega_{c}(\mathbf{r})
\end{equation}

where $\Omega_{f}(\mathbf{r})$ is a shape function of the whole fin, $V_{o}$ - constant potential of Hafnium Oxide,
$\Omega_{c}(\mathbf{r})$ - shape function of the core, $V_{c}$ - constant potential of the core.
In the first part of equation we create a hollow rectangular fin by extracting from the whole Hafnium Oxide fin its core part
and then adding the core shape again but with silicon potential $V_c$.
In \emph{ESCAPE} each term in equation \ref{eq:fin_potential} correspond to potential object as following

\begin{python}
V_of = esc.box("", qx, qy, qz, hfo2,
               wx_f, wy_f, wz_f,
               pos=[(-D/2, 0, wz_f/2),
                    (D/2, 0, wz_f/2)])
V_oc = esc.box("", qx, qy, qz, hfo2,
                wx_c, wy_c, wz_c,
                pos=[(-D/2, 0, wz_c/2),
                     (D/2, 0, wz_c/2)])
V_cc = esc.box("", qx, qy, qz, si,
               wx_si, wy_c, wz_c,
               pos=[(-D/2.0, 0, wz_c/2),
                    (D/2.0, 0, wz_c/2)])
V_hol = V_of - V_oc
V = V_hol + V_cc

show(V_hol, title="Hollow fin")
show(V, title="Filled fin")
\end{python}

where $V_{hol}$ is a potential of the hollow rectangular fin and $V$ is a final potential. Seemingly the presented
potential arithmetic operations are similar to real processes used in lithography, like etching or thin film deposition.
Even for such simple structures, which we presented here, it is already not the case.
The purpose of these operations is to describe a simulated structure, but not a physical process of their preparation.
The potential object calculates form-factor of resulted shape using divergence theorem. 
The formulas for the form-factor of 3D polyhedra were first presented by Liu, (2011),
based on work of McInturff et al. (1991) for 2D polygons.

\begin{figure}
\includegraphics[width=\textwidth]{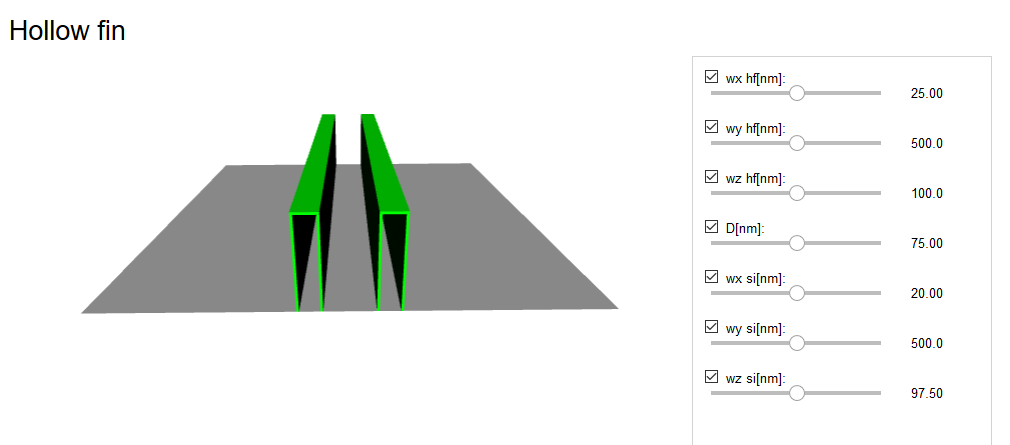}
\caption{Widget for potential shape. Example with hollow fin.}
\label{fig:widget3}
\end{figure}

\begin{figure}
\includegraphics[width=\textwidth]{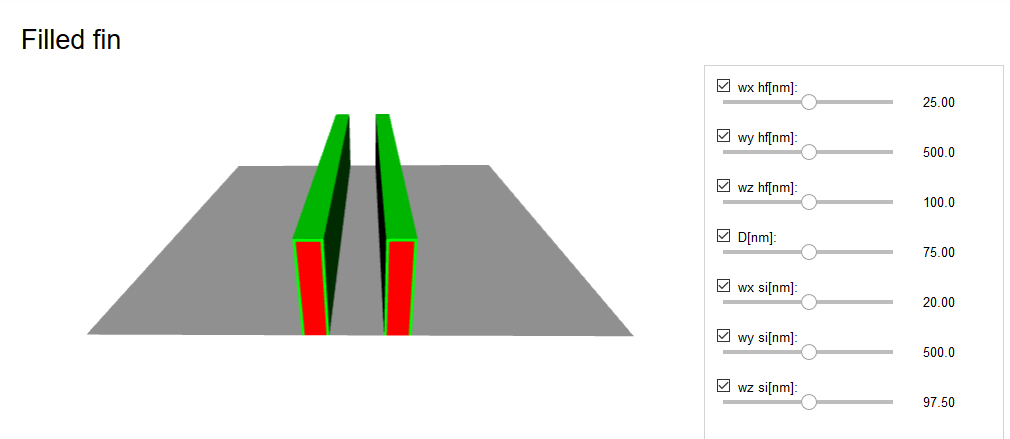}
\caption{Widget for potential shape. Example with filled fin.}
\label{fig:widget4}
\end{figure}

\subsubsection{Structured samples. Small Angle Scattering}

In previous section we described by means of potential our structure unit which consist of two Si fins covered with hafnium oxide.
In this section we are going to use the created potential object for calculation of three-dimensional Small Angle Scattering
pattern from a regular one-dimensional lattice of such fins. For this purpose we add a description of the lattice as following

\begin{python}
T=esc.par("Period X", 200,  units="nm")
latx = esc.lattice("", qx, T, 5)
show(latx, coordinates=np.linspace(-0.1, 0.1, 500),
     xlabel="Qx")
\end{python}

\begin{figure}
\includegraphics[width=\textwidth]{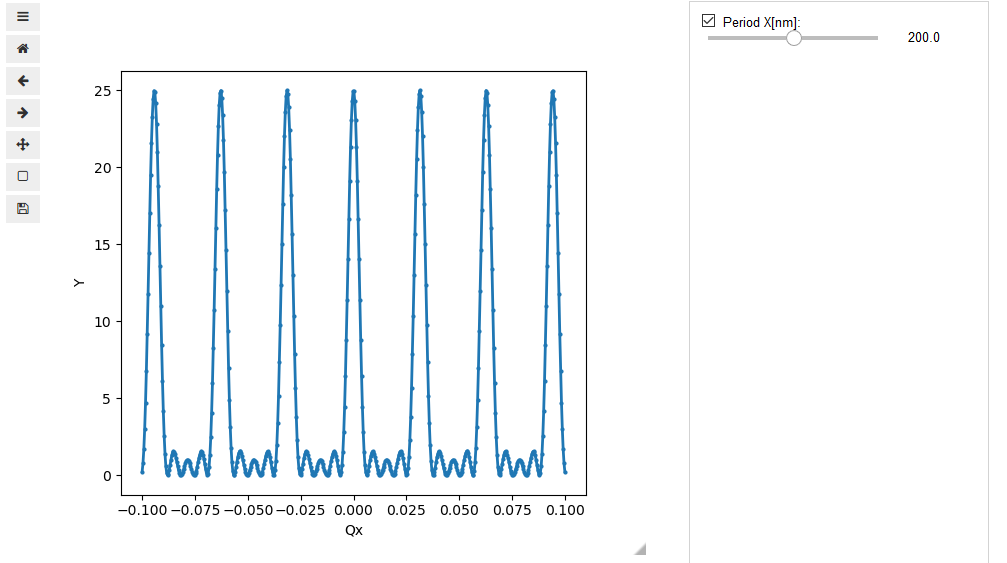}
\caption{Widget for lattice functor.}
\label{fig:widget5}
\end{figure}

The \emph{latx} object is a one-dimensional functor which returns result for the lattice function of the following form
\begin{equation}
\label{eq:lattice}
L(q_x)=\left|\frac{\sin(N q_x T 0.5)}{\sin(q_x T 0.5)}\right|^2
\end{equation}

where $N$ - number of structure units in the lattice, which we have set as $N=5$.
The resulted functor is depicted in figure \ref{fig:widget5}

Since \emph{latx} is a functor, if equation \ref{eq:lattice} doesn't satisfy the current needs, any other custom functor can be used instead.
The two-dimensional lattice is also quite simple to implement, user has to create the second lattice functor along another axis, for example
$q_y$, and simply multiply both \pyth{latx*laty}. The resulted functor correspond to the two-dimensional lattice along $q_x$ and $q_y$ axes.

In the next code listing we demonstrate how to create a functor which calculates Small Angle Scattering intensity. In addition to three variables,
which correspond to scattering vector components $q_x,\;q_y,\;q_z$, it requires potential object and lattice as a functor of one or several
variables.

\begin{python}
I = esc.sas("", qx, qy, qz, V, latx)
show(I, coordinates=c, cblog=True,
     cbmin=1, rows=sz,
     plot_type="map", xlabel="Qx", ylabel="Qy")
\end{python}

\begin{figure}
\includegraphics[width=\textwidth]{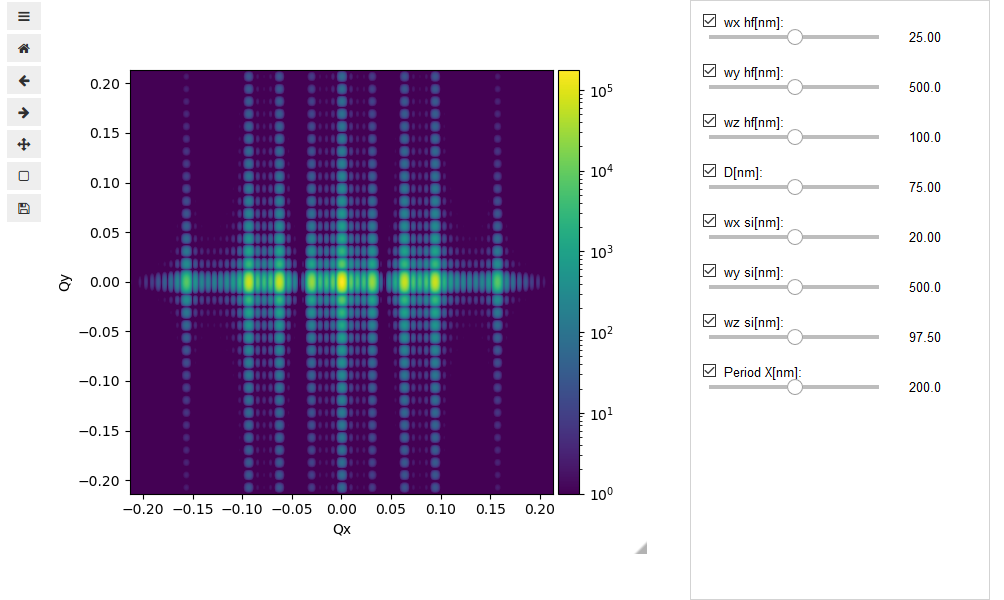}
\caption{Calculated small angle scattering pattern.}
\label{fig:widget6}
\end{figure}

The result of \emph{show} command is depicted in figure \ref{fig:widget6}.

\section{Conclusion and outlook}

In this article we presented the key features of the first public release, version 0.9.0, of \emph{ESCAPE} software project.
It took quite a lot of time to analyse advantages and disadvantages of existing scattering software, including
commercial packages, and to find the best architecture for the core suitable for most of the scattering applications,
balancing between experienced and non-experienced users.
We believe that our way of software design, way of working with the data analysis will find followers in the
scattering community. The standard format of Jupyter notebooks should help users of \emph{ESCAPE} to share their experience with others
using our repository (check www.escape-app.net). Below we give an outlook of the future features of \emph{ESCAPE}.

\subsection{Potential module and structured samples}

Right now we are concentrated on further development of \emph{potential} module, which we briefly described in section \ref{sec:potential}.
The \emph{potential} object supports simple arithmetic operations, allowing to create complicated shapes, forming a structure unit
and to position them, defining their coordinates as independent fit parameters. Our users can operate with predefined existing shapes,
use \emph{geometry} object for creation of complex shapes based on simple shapes, or describe fully custom shape providing vertices coordinates.
Each component of vertices coordinates is a parameter and user can easily link them to physical parameters of the described shape if necessary.

Currently, there is no control for shape overlapping, it is up to the user to check the correctness of parameters boundaries.
We plan to add overlapping control in the next release together with the support for structured layers, necessary for GISAS and off-specular reflectivity
applications.

\subsection{Off-specular applications}

The \emph{potential} module will be the basis for new off-specular scattering functors.
One can split the off-specular scattering applications on two branches: grazing incidence cases, like grazing incidence small angle scattering,
grazing incidence diffraction, off-specular reflectometry and non-grazing cases, like 2D intensity maps of high resolution X-ray diffraction.
The former types are not very popular in industrial applications due to the large footprint of an incident beam on the sample, but the latter one
is used intensively for structured samples.
The basis for these applications is already implemented in HRXRD and specular reflectivity modules. The objects for sample description
for these cases do not require any significant update.
We plan to implement them using kinematic, semi-kinematic and dynamic approximations.

\subsection{Small angle scattering. Diluted samples}

It is currently unclear for us whether we need a special module for small angle scattering for diluted samples.
In our examples we have demonstrated that experiments can be simulated using only the \emph{core} package.
The development of this branch could  go into two directions: adding more example notebooks covering most of
the usual form-factors and user cases or adding these typical form-factors to the C++ core, which will certainly make
calculation faster, but will reduce possibilities for customization.
A more complicated cases could arise from the particles of complex shapes where the analytical expression of the
form-factor can be quite difficult to achieve. Here the \emph{potential} and \emph{geometry} modules could help, especially
if the same probes have been measured with conventional small angle scattering and GISAS.
In this case the characteristic parameters of the particles could be shared between models and simultaneous fit performed.
We leave this decision for the future community right now.

\subsection{Acknowledgements}

We thank Yury Khaydukov for providing experimental data for testing and performing first in situ tests and
Miguel Castro-Collin for early testing, discussions and for his helpful feedback.
Special thanks to Oleg Khoruzhiy for discussion about neural networks and useful references.





\end{document}